\documentclass[aps,prl,reprint,nobibnotes,nofootinbib]{revtex4-1}
\usepackage{amsmath,amssymb,amsfonts,amsthm,mathtools}
\usepackage{bbm}
\usepackage{hyperref}
\hypersetup{colorlinks=true,linkcolor=blue,citecolor=blue,urlcolor=blue}
\usepackage{slashed} 
\usepackage{braket} 
\usepackage{simplewick} 
\newtheorem{theorem}{Theorem}[section]

\newtheorem{definition}[theorem]{Definition}

\newcommand{\R}{\mathbb{R}}
\newcommand{\C}{\mathbb{C}}

\newcommand{\dd}{\mathrm{d}}
\newcommand{\ii}{\mathrm{i}}

\usepackage{graphicx}
\usepackage{tikz}
\usepackage{tikz-feynman}
\tikzfeynmanset{compat=1.1.0}

\begin{document}

\title{On the Four-Loop Higgs--Gluon Form Factor in Nonlocal Quantum Field Theory}
\author{E.~J.~Thompson}
\affiliation{Wilfrid Laurier University, Waterloo, Canada, N2L 3C5}
\email{thom3471@mylaurier.ca}
\date{\today}

\begin{abstract}
In particle physics the virtual N$^3$LO contribution to gluon-fusion Higgs production in the Standard Model requires four-loop multi-scale QCD amplitudes that in dimensional regularization are dominated by singular integral reduction and difficult master-integral evaluations. In this note we show that a nonlocal UV completion makes each fixed-order four-loop Feynman diagram fully convergent in four dimensions while keeping the analytic structure needed for Lehmann–Symanzik–Zimmermann (LSZ) reduction formula. For gluon-fusion $gg\!\to\!H$ we derive a closed form Schwinger-parameter representation where all loop-momentum integrals are performed analytically and thus leaving finite parameter integrals viable for direct numerical evaluation without needing the usual UV subtractions. This method heps to isolate the collider-hard contribution as a convergent four-loop form factor. The strong coupling then is defined by a finite normalization condition in the entire-function regularization scheme and it is related to the $\overline{\mathrm{MS}}$ coupling by finite matching where after this matching, the regulator-dependent corrections decouple in the local limit $E_M\!\to\!\infty$.
\end{abstract}

\maketitle

\section{Introduction}

The usual computational bottleneck in precision collider theory is not just the definition of some observables, but it is the precise solution of the multi-loop field equations from Feynman integrals where integration-by-parts reduction produces enormous linear systems and where the resulting master integrals often satisfy stiff coupled differential equations in several mass ratios, this makes them extremity difficult to work out realistically~~\cite{tHooftVeltman1972,BolliniGiambiagi1972,ChetyrkinTkachov1981,Laporta2000,Kotikov1991,Remiddi1997,Henn2013,LSZ1955,Cutkosky1960,F0Normalization,ChetyrkinTkachov1981,Laporta2000,Kotikov1991,Remiddi1997,Henn2013,BognerWeinzierl2010}. A "simple" example we can work out is the virtual N$^3$LO contribution to gluon-fusion Higgs production. In which the Standard Model leading-order amplitude already begins at one loop and the N$^3$LO virtual correction requires four-loop QCD corrections to the $ggH$ form factor~\cite{Dawson1991,Anastasiou2015N3LOthreshold,Mistlberger2018N3LO,Baikov2009FormFactors,Gehrmann2010eps2,LeeSmirnovSmirnovSteinhauser2022}. Our goal here is to show that an entire-function regulator from nonlocal quantum field theory converts these gross UV-sensitive loop equations into absolutely convergent integrals in four dimensions, and we will give an explicit four-loop representation where all loop-momentum integrations can be carried out in closed form. Ultraviolet finiteness does not fix the finite normalization of the QCD coupling, so below we define the coupling in the entire-function scheme and give its finite matching relation to the standard $\overline{\mathrm{MS}}$ coupling.

The present construction should be seen as the non-Abelian Higgs-sector analogue of the asymptotic renormalization program previously developed for regulated QED~\cite{Thompson2026AR}.

\section{THE ENTIRE-FUNCTION COMPLETION OF QCD}

We start from the standard QCD and top-Yukawa sector in Minkowski space~\cite{Moffat2025EWSB} and implement the UV completion by analytic dressing of kinetic operators with an entire function of a gauge-covariant Laplace--Beltrami operator in the background-field formalism~\cite{Abbott1982,Moffat1990,Evens1991,MoffatThompsonGaugeEntire2026,ThompsonCovariance2026}. For the calculation we do here the only properties we require are analyticity and strong Euclidean damping~\cite{Moffat1990,Evens1991,MoffatThompsonGaugeEntire2026,ModestoRachwal2015,ModestoPivaRachwal2016}. We define the desired properties of our entire function as:

\begin{definition}[Admissible entire regulator] 
Let $E_M>0$ be the intrinsic UV scale and let $F:\C\to\C$ be entire.
We call $F$ admissible if
\begin{align}
F(0)&=1,
\\
|F(-p_E^2/E_M^2)|&\le \exp(-c\,p_E^2/E_M^2),
\label{eq:admissibleF}
\end{align}
for $p_E^2\gg E_M^2$, and for some constant $c>0$.
\end{definition}
\noindent Here $E_M$ is the Moffat parameter and is defined as the energy in which in QFT the nonlocality must take into account, and $p_E\in\R^4$ is Euclidean momentum and: 
\begin{equation}
p_E^2=\sum_{j=1}^4(p_E^j)^2,
\end{equation}
is its squared norm, while $c$ controls the UV damping rate.

In momentum space each internal propagator in a fixed-order expansion acquires a multiplicative factor of $F(-\ell^2/E_M^2)$, where $\ell$ is the momentum flowing through that internal line. For a massive fermion line, like for the top quark we use the schematic regulated propagator:
\begin{equation}
S_F^{(F)}(\ell)=
\frac{\ii(\slashed{\ell}+m_t)\,F(-\ell^2/E_M^2)}
{\ell^2-m_t^2+\ii 0},
\label{eq:fermprop}
\end{equation}
where $m_t$ is the top mass and $\slashed{\ell}=\gamma^\mu \ell_\mu$, meaning you turn a 4-vector into a Dirac-matrix–valued object by dotting it into the gamma matrices that are the matrix representation of the Clifford algebra associated with spacetime. For a gluon line in a covariant gauge we use:
\begin{equation}
D_{\mu\nu}^{ab\,(F)}(\ell)=
\frac{\ii\,\delta^{ab}\,F(-\ell^2/E_M^2)}
{\ell^2+\ii 0}
\left(-g_{\mu\nu}+(1-\xi)\frac{\ell_\mu\ell_\nu}{\ell^2+\ii 0}\right),
\label{eq:gluonprop}
\end{equation}
where $\xi$ is the gauge parameter, $a,b$ are adjoint colour indices, and $g_{\mu\nu}$ is the Minkowski metric. Equations \eqref{eq:fermprop} and \eqref{eq:gluonprop} display only the quadratic-sector propagator dressing, so we note that in the fully gauge-completed theory, the interaction vertices are obtained by expanding the same gauge-covariant entire operators that appear in the action. This expansion generates correlated momentum-dependent quark--gluon, three-gluon, four-gluon, ghost, and Yukawa vertices, together with any measure contribution required by the chosen BRST-invariant quantization prescription~\cite{Abbott1982,Moffat1990,Evens1991,MoffatThompsonGaugeEntire2026}. The amplitudes and self-energies below are understood to be computed using the complete gauge-related set of regulated Feynman rules, not by dressing propagators while leaving all interaction vertices unchanged. The background-field Ward--Slavnov--Taylor identities then relate the normalization of the gauge coupling to the background-gluon Green functions.

Although the nonlocal completion makes the relevant loop integrals ultraviolet finite, the finiteness alone does not determine the finite normalization of the QCD coupling. We thus need to distinguish the coupling defined in the entire-function scheme from the conventional $\overline{\mathrm{MS}}$ coupling. We can do this by writing:
\begin{align}
\alpha_s^{(F)}(\mu,E_M)
&\equiv
\frac{\bigl[g_s^{(F)}(\mu,E_M)\bigr]^2}{4\pi},
\\
a_F(\mu,E_M)
&\equiv
\frac{\alpha_s^{(F)}(\mu,E_M)}{4\pi},
\label{eq:aFdefinition}
\end{align}
where $\mu$ is the finite coupling-normalization scale.

A gauge-invariant definition is obtained in the background-field formalism, so we let:
\begin{equation}
\Gamma_{\bar A\bar A\bar A}^{abc,\mu\nu\rho}(p_1,p_2,p_3),
\end{equation}
represent the renormalized one-particle-irreducible background three-gluon vertex, and let $\mathcal{T}_{F,0}^{abc,\mu\nu\rho}(p_1,p_2,p_3)$ be its gauge-completed tree-level tensor with the coupling removed. We can then define $g_s^{(F)}(\mu,E_M)$ by the Euclidean symmetric-point condition:
\begin{equation}
\left.
\Gamma_{\bar A\bar A\bar A}^{abc,\mu\nu\rho}
(p_1,p_2,p_3)
\right|_{\mathrm{sym}}
=
g_s^{(F)}(\mu,E_M)
\left.
\mathcal{T}_{F,0}^{abc,\mu\nu\rho}
(p_1,p_2,p_3)
\right|_{\mathrm{sym}},
\label{eq:Fcouplingcondition}
\end{equation}
where:
\begin{align}
p_{1E}^2&=p_{2E}^2=p_{3E}^2=\mu^2,
\\
p_{iE}\!\cdot p_{jE}&=-\frac{\mu^2}{2}
\quad (i\neq j),
\\
p_1+p_2+p_3&=0.
\label{eq:symmetricpoint}
\end{align}
And by the background-field Ward identities the same coupling can equivalently be extracted from the transverse background-gluon two-point function~\cite{Abbott1982}. All loop integrals entering this condition are finite in the nonlocal theory, but their finite parts depend on $E_M$ and on the normalization condition \eqref{eq:Fcouplingcondition}.

The coupling $\alpha_s^{(F)}$ is not assumed to be identical to $\alpha_s^{\overline{\mathrm{MS}}}$ at finite $E_M$. Their relation is a finite perturbative matching relation:
\begin{align}
a_F(\mu,E_M)
={}&
a_{\overline{\mathrm{MS}}}(\mu)
+c_1\!\left(\frac{\mu^2}{E_M^2}\right)
a_{\overline{\mathrm{MS}}}^2(\mu)
\nonumber\\
&+
c_2\!\left(\frac{\mu^2}{E_M^2}\right)
a_{\overline{\mathrm{MS}}}^3(\mu)
+\mathcal{O}\!\left(a_{\overline{\mathrm{MS}}}^4\right),
\label{eq:couplingmatching}
\end{align}
where:
\begin{equation}
a_{\overline{\mathrm{MS}}}(\mu)
\equiv
\frac{\alpha_s^{\overline{\mathrm{MS}}}(\mu)}{4\pi}.
\end{equation}
The finite coefficients $c_1$ and $c_2$ are determined by evaluating the same gauge-invariant reference Green function in the two prescriptions and requiring that:
\begin{equation}
\left.
\Gamma_{\bar A\bar A\bar A}^{(F)}
\bigl(a_F,E_M\bigr)
\right|_{\mathrm{sym}}
=
\left.
\Gamma_{\bar A\bar A\bar A}^{\overline{\mathrm{MS}}}
\bigl(a_{\overline{\mathrm{MS}}},\mu\bigr)
\right|_{\mathrm{sym}}
\label{eq:matchingcondition}
\end{equation}
order by order in the coupling. So then $c_1$ is fixed by the finite one-loop difference and $c_2$ by the finite two-loop difference. So there should be no assumption equality of the two couplings before this matching is performed.

Ultraviolet finiteness also does not imply a vanishing beta function, we note that the running coupling in the entire-function scheme is defined by:
\begin{equation}
\beta_F(a_F,\mu/E_M)
\equiv
\left.
\mu^2\frac{\dd a_F}{\dd\mu^2}
\right|_{\mathrm{bare},\,E_M},
\label{eq:betaF}
\end{equation}
and follows from differentiating the finite matching relation
\eqref{eq:couplingmatching}. In the limit $E_M\to\infty$, the coefficients $c_i$ approach the usual finite conversion coefficients between the chosen symmetric-point normalization and the $\overline{\mathrm{MS}}$ scheme. The physical local-QCD result is recovered after this finite scheme conversion where the two scheme-dependent couplings need not be numerically identical without an explicit matching convention.

\section{THE GLUON-FUSION HIGGS FORM FACTOR}
We consider gluon fusion to a Higgs, $g(p_1,a,\mu)+g(p_2,b,\nu)\to H(q)$ with:
\begin{equation}
q=p_1+p_2,
\qquad
p_1^2=p_2^2=0,
\qquad
q^2=m_H^2,
\label{eq:kinematics}
\end{equation}
where $m_H$ is the mass of the Higgs and $\mu,\nu$ are Lorentz indices of the external gluons. Gauge invariance says that the amplitude can be written as a single scalar form factor multiplying the transverse tensor structure~\cite{Dawson1991,Baikov2009FormFactors,Gehrmann2010eps2,LeeSmirnovSmirnovSteinhauser2022}:
\begin{align}
\notag \mathcal{M}_{\mu\nu}^{ab}(p_1,p_2)=
\delta^{ab}\,
\left(p_1\!\cdot\! p_2\,g_{\mu\nu}-p_{2\mu}p_{1\nu}\right)\,
\\\times\mathcal{F}_{ggH}(m_H^2,m_t^2;E_M^2),
\label{eq:FFdecomp}
\end{align}
where $\mathcal{F}_{ggH}$ is the regulated $ggH$ form factor. The dot means that $p_1\!\cdot\! p_2=g_{\rho\sigma}p_1^\rho p_2^\sigma$.

We will first expand the scalar form factor in terms of the coupling defined by the entire-function normalization condition of \eqref{eq:Fcouplingcondition}:
\begin{align}
\notag
\mathcal{F}_{ggH}
={}&
\mathcal{F}^{(0)}_F
+a_F(\mu,E_M)\mathcal{F}^{(1)}_F
+a_F^2(\mu,E_M)\mathcal{F}^{(2)}_F
\\
&+
a_F^3(\mu,E_M)\mathcal{F}^{(3)}_F
+\mathcal{O}(a_F^4),
\label{eq:alphasexpansion}
\end{align}
where $\mathcal{F}^{(0)}_F$ is the one-loop Standard Model triangle contribution and $\mathcal{F}^{(3)}_F$ contains the four-loop N$^3$LO virtual contribution~\cite{Dawson1991,Baikov2009FormFactors,Gehrmann2010eps2, LeeSmirnovSmirnovSteinhauser2022}. The subscript $F$ denotes coefficients defined using the entire-function coupling and its finite normalization condition. In general:
\begin{equation}
\mathcal{F}^{(n)}_F
=
\mathcal{F}^{(n)}_F
(m_H^2,m_t^2;\mu^2,E_M^2).
\end{equation}

For comparison with phenomenological results quoted in the $\overline{\mathrm{MS}}$ scheme we would have to substitute the matching relation \eqref{eq:couplingmatching} into \eqref{eq:alphasexpansion}. By writing:
\begin{equation}
\mathcal{F}_{ggH}
=
\mathcal{F}^{(0)}_{\overline{\mathrm{MS}}}
+
a_{\overline{\mathrm{MS}}}
\mathcal{F}^{(1)}_{\overline{\mathrm{MS}}}
+
a_{\overline{\mathrm{MS}}}^2
\mathcal{F}^{(2)}_{\overline{\mathrm{MS}}}
+
a_{\overline{\mathrm{MS}}}^3
\mathcal{F}^{(3)}_{\overline{\mathrm{MS}}}
+\cdots,
\label{eq:MSbarexpansion}
\end{equation}
the strong-coupling conversion gives:
\begin{align}
\mathcal{F}^{(0)}_{\overline{\mathrm{MS}}}
&=
\mathcal{F}^{(0)}_F,
\\
\mathcal{F}^{(1)}_{\overline{\mathrm{MS}}}
&=
\mathcal{F}^{(1)}_F,
\\
\mathcal{F}^{(2)}_{\overline{\mathrm{MS}}}
&=
\mathcal{F}^{(2)}_F
+
c_1\mathcal{F}^{(1)}_F,
\\
\mathcal{F}^{(3)}_{\overline{\mathrm{MS}}}
&=
\mathcal{F}^{(3)}_F
+
2c_1\mathcal{F}^{(2)}_F
+
c_2\mathcal{F}^{(1)}_F.
\label{eq:coefficientconversion}
\end{align}
The arguments $\mu^2/E_M^2$ of $c_1$ and $c_2$ are suppressed in \eqref{eq:coefficientconversion}. Equation \eqref{eq:coefficientconversion} shows that ultraviolet finiteness of the individual integrals does not make their perturbative coefficients scheme independent. Only the complete form factor after all parameters have been expressed in one common scheme is a physical quantity.

\section{The explicit four-loop reduction to finite parameter integrals}

The new N$^3$LO loop object is the connected four-loop contribution computed in the entire-function scheme. We represent this raw four-loop coefficient by:
\begin{equation}
\mathcal{F}^{(3)}_{F,\mathrm{4L}}
(m_H^2,m_t^2;E_M^2)
=
\sum_{G\in\mathcal{G}_4}
C_G\,
I_G(m_H^2,m_t^2;E_M^2),
\label{eq:F3sum}
\end{equation}
where $C_G$ are colour factors and $I_G$ are scalarized projected integrals obtained after tensor reduction at the integrand level. The important point is that the integrals $I_G$ are absolutely convergent in four dimensions so long as $F$ satisfies \eqref{eq:admissibleF}.

The coefficient entering an expansion in a specified renormalized coupling must also contain the corresponding finite normalization and matching contributions, so in particular after conversion to $\alpha_s^{\overline{\mathrm{MS}}}$, the strong-coupling part of the N$^3$LO coefficient is given by:
\begin{equation}
\mathcal{F}^{(3)}_{\overline{\mathrm{MS}}}
=
\mathcal{F}^{(3)}_{F,\mathrm{4L}}
+
2c_1\mathcal{F}^{(2)}_F
+
c_2\mathcal{F}^{(1)}_F,
\label{eq:F3matched}
\end{equation}
up to any separately chosen finite conversions of the top mass, Yukawa coupling, or external-field normalization. Equation \eqref{eq:F3sum} then therefore isolates the four-loop integral contribution, while Eq.~\eqref{eq:F3matched} gives the coupling-matched coefficient required for comparison with a conventional $\overline{\mathrm{MS}}$ prediction.

We can illustrate the core mechanism of this model on a generic four-loop integral with internal line momenta $\ell_r(k,p)$, where $k=(k_1,k_2,k_3,k_4)$ are independent loop momenta and $p=(p_1,p_2)$ are external. A typical scalarized integrand has the form of:
\begin{align}
\notag I_G=
\int\prod_{a=1}^4\frac{\dd^4 k_a}{(2\pi)^4}\;
\frac{N_G(k,p)}{\prod_{r=1}^{L_G}\left(\ell_r(k,p)^2-m_r^2+\ii 0\right)^{\nu_r}}\;
\\\times\prod_{r=1}^{L_G}F\!\left(-\frac{\ell_r(k,p)^2}{E_M^2}\right),
\label{eq:generic4loop}
\end{align}
where $L_G$ is the number of internal lines, $m_r\in\{0,m_t\}$ are internal masses, $\nu_r\in\mathbb{N}$ are propagator powers, and $N_G$ is a polynomial numerator from Dirac traces and Lorentz contractions.

We now Wick rotate each $k_a^0\mapsto \ii k_{aE}^4$ so $\ell_r^2\mapsto -\ell_{rE}^2$.
Under the admissibility conditions of \eqref{eq:admissibleF}, the factors $F(-\ell_r^2/E_M^2)$ become exponentially damping functions of $\ell_{rE}^2/E_M^2$, and the integral is dominated by a Gaussian tail at large Euclidean loop momentum.

We introduce Schwinger parameters for each denominator~\cite{Schwinger1951}:
\begin{equation}
\frac{1}{(\ell_{rE}^2+m_r^2)^{\nu_r}}
=
\frac{1}{\Gamma(\nu_r)}
\int_0^\infty \dd\alpha_r\;\alpha_r^{\nu_r-1}\,
\exp\!\left[-\alpha_r(\ell_{rE}^2+m_r^2)\right],
\label{eq:schwinger}
\end{equation}
where $\Gamma$ is the Euler gamma function and $\alpha_r>0$ is the Schwinger parameter associated with line $r$.

For a simple implementable choice capturing the entire-function class, we take the representative regulator~\cite{MoffatThompsonGaugeEntire2026,ModestoRachwal2015,ModestoPivaRachwal2016}:
\begin{equation}
F(z)=\exp(z),
\quad
\Rightarrow\quad
F\!\left(-\frac{\ell_{rE}^2}{E_M^2}\right)=\exp\!\left(-\frac{\ell_{rE}^2}{E_M^2}\right),
\label{eq:regchoice}
\end{equation}
so each internal line contributes an extra Gaussian factor in Euclidean space. Combining \eqref{eq:schwinger} with \eqref{eq:regchoice} gives us, line by line:
\begin{align}
\notag \exp\!\left[-\alpha_r(\ell_{rE}^2+m_r^2)\right]\;
\exp\!\left(-\frac{\ell_{rE}^2}{E_M^2}\right)
\\=
\exp\!\left[-\Big(\alpha_r+\frac{1}{E_M^2}\Big)\ell_{rE}^2-\alpha_r m_r^2\right].
\label{eq:combinealpha}
\end{align}
So the regulator is absorbed as a positive shift $\alpha_r\mapsto \alpha_r+E_M^{-2}$ in the quadratic form controlling the Euclidean loop-momentum Gaussian.

After introducing all Schwinger parameters, the Euclidean integrand becomes a polynomial in $k_{aE}$ times a single exponential of a quadratic form in the $16$ components of $(k_{1E},k_{2E},k_{3E},k_{4E})$. We now introduce a $16$-component column vector $K$ collecting all Euclidean loop components:
\begin{widetext}
\begin{equation}
\begin{aligned}
\begin{split}
K\equiv (k_{1E}^1,\ldots,k_{1E}^4;\;k_{2E}^1,\ldots,k_{2E}^4;\;k_{3E}^1,\ldots,k_{3E}^4;\;k_{4E}^1,\ldots,k_{4E}^4)^T.
\label{eq:Kdef}
\end{split}
\end{aligned}
\end{equation}
\end{widetext}
Then the exponent can be written as:
\begin{align}
\notag \sum_{r=1}^{L_G}\Big(\alpha_r+\frac{1}{E_M^2}\Big)\ell_{rE}^2
=
&K^T A(\alpha;E_M) K \\&- 2\,B(\alpha;p)^T K + C(\alpha;p),
\label{eq:quadform}
\end{align}
where $A$ is a $16\times 16$ positive-definite matrix, $B$ is a $16$-vector linear in external momenta, and $C$ is a scalar depending on external invariants and masses. Here we have $\alpha=(\alpha_1,\ldots,\alpha_{L_G})$, meaning the collection of Schwinger/Feynman parameters, one for each internal line propagator of the graph $G$. Completing the square through $K\mapsto K-A^{-1}B$ gives us:
\begin{equation}
K^T A K-2B^T K
=
(K-A^{-1}B)^T A (K-A^{-1}B)-B^T A^{-1}B.
\label{eq:completesquare}
\end{equation}
Since $A$ is positive definite the full $K$-integral is Gaussian and can be performed analytically. For a scalar numerator $N_G=1$ the result is:
\begin{widetext}
\begin{equation}
\begin{aligned}
\begin{split}
\int\frac{\dd^{16}K}{(2\pi)^{16}}\;
\exp\!\left[-(K-A^{-1}B)^T A (K-A^{-1}B)\right]
=
\frac{1}{(4\pi)^8}\frac{1}{\sqrt{\det A(\alpha;E_M)}},
\label{eq:gaussian}
\end{split}
\end{aligned}
\end{equation}
\end{widetext}
where $\det A$ is the determinant of the $16\times 16$ matrix. For polynomial numerators $N_G(k,p)$ we introduce a source componant $J$:
\begin{align}
Z(J)&\equiv
\int\frac{\dd^{16}K}{(2\pi)^{16}}\;
\exp\!\left[-K^T A K+2J^T K\right]
\\&=
\frac{1}{(4\pi)^8}\frac{1}{\sqrt{\det A}}\exp\!\left(J^T A^{-1}J\right),
\label{eq:source}
\end{align}
and then we represent $K$-moments by derivatives $\partial/\partial J$ evaluated at $J=B(\alpha;p)$. This yields a purely algebraic map from $N_G$ to contractions of $A^{-1}$ with external momenta.

Putting the pieces together we see that each graph integral takes the explicit Schwinger-parameter form:
\begin{widetext}
\begin{equation}
\begin{aligned}
\begin{split}
I_G=
\frac{1}{(4\pi)^8}
\left[\prod_{r=1}^{L_G}\frac{1}{\Gamma(\nu_r)}\int_0^\infty \dd\alpha_r\;\alpha_r^{\nu_r-1}\right]
\frac{\mathcal{P}_G(\alpha;p,m_t)}{\sqrt{\det A(\alpha;E_M)}}\;
\exp\!\left[-\Phi_G(\alpha;p,m_t;E_M)\right],
\label{eq:IGfinal}
\end{split}
\end{aligned}
\end{equation}
\end{widetext}
where $\mathcal{P}_G$ is a polynomial formed from $A^{-1}$ contractions encoding the original numerator, and $\Phi_G$ is the positive exponent:
\begin{widetext}
\begin{equation}
\begin{aligned}
\begin{split}
\Phi_G(\alpha;p,m_t;E_M)\equiv
\sum_{r=1}^{L_G}\alpha_r m_r^2
+ C(\alpha;p) - B(\alpha;p)^T A(\alpha;E_M)^{-1}B(\alpha;p).
\label{eq:Phi}
\end{split}
\end{aligned}
\end{equation}
\end{widetext}
At the level of each raw four-loop graph, the dependence on the UV completion enters through the shifted quadratic form in $A(\alpha;E_M)$ and therefore through $\det A$ and $A^{-1}$. The renormalized perturbative coefficient also contains the finite coupling-normalization and scheme-matching terms displayed in Eqs.~\eqref{eq:couplingmatching} and \eqref{eq:F3matched}.

It may be useful to display a fully explicit lower-loop prototype of the same mechanism that underlies the four-loop representation in Eq.~\eqref{eq:IGfinal} to show exactly what we are doing and why it is useful. If we consider the scalarized Euclidean one-loop two-propagator integral:
\begin{widetext}
\begin{equation}
\begin{aligned}
\begin{split}
I_{2}(q_E^2;m_t^2,E_M^2)
=
\int \frac{d^4 k_E}{(2\pi)^4}
\frac{\exp(-k_E^2/E_M^2)\,\exp(-(k_E+q_E)^2/E_M^2)}
{(k_E^2+m_t^2)\big((k_E+q_E)^2+m_t^2\big)},
\label{eq:workedprototype}
\end{split}
\end{aligned}
\end{equation}
\end{widetext}
we note that this is not yet the full one-loop $ggH$ triangle, but it is the minimal nontrivial example showing simply how the nonlocal regulator takes an ultraviolet-divergent Feynman integral into a finite parameter integral of exactly the same type as the general four-loop formula.

We start by introducing Schwinger parameters:
\begin{align}
\frac{1}{k_E^2+m_t^2}
&=
\int_0^\infty d\alpha\,
e^{-\alpha(k_E^2+m_t^2)},
\\
\frac{1}{(k_E+q_E)^2+m_t^2}
&=
\int_0^\infty d\beta\,
e^{-\beta((k_E+q_E)^2+m_t^2)},
\end{align}
we then obtain:
\begin{widetext}
\begin{equation}
\begin{aligned}
\begin{split}
I_{2}
&=
\int_0^\infty d\alpha \int_0^\infty d\beta
\int \frac{d^4 k_E}{(2\pi)^4}
\exp\Big[
-(\alpha+E_M^{-2})k_E^2
-(\beta+E_M^{-2})(k_E+q_E)^2
-(\alpha+\beta)m_t^2
\Big].
\end{split}
\end{aligned}
\end{equation}
\end{widetext}
Then expanding the second quadratic form gives us:
\begin{widetext}
\begin{equation}
\begin{aligned}
\begin{split}
I_{2}
&=
\int_0^\infty d\alpha \int_0^\infty d\beta
\int \frac{d^4 k_E}{(2\pi)^4}
\exp\Big[
-A\,k_E^2
-2(\beta+E_M^{-2})\,k_E\!\cdot\! q_E
-(\beta+E_M^{-2})q_E^2
-(\alpha+\beta)m_t^2
\Big],
\end{split}
\end{aligned}
\end{equation}
\end{widetext}
with:
\begin{equation}
A \equiv \alpha+\beta+2E_M^{-2} >0.
\end{equation}
Then completing the square:
\begin{widetext}
\begin{equation}
\begin{aligned}
\begin{split}
A k_E^2 + 2(\beta+E_M^{-2})k_E\!\cdot\! q_E
=
A\left(k_E+\frac{\beta+E_M^{-2}}{A}q_E\right)^2
-
\frac{(\beta+E_M^{-2})^2}{A}q_E^2,
\end{split}
\end{aligned}
\end{equation}
\end{widetext}
so the exponent becomes:
\begin{widetext}
\begin{equation}
\begin{aligned}
\begin{split}
-A\left(k_E+\frac{\beta+E_M^{-2}}{A}q_E\right)^2
-
(\alpha+\beta)m_t^2
-
\frac{(\alpha+E_M^{-2})(\beta+E_M^{-2})}{A}\,q_E^2.
\end{split}
\end{aligned}
\end{equation}
\end{widetext}
The Gaussian $k_E$-integral is now elementary:
\begin{equation}
\int \frac{d^4 k_E}{(2\pi)^4} e^{-A k_E^2}
=
\frac{1}{(4\pi)^2}\frac{1}{A^2},
\end{equation}
\clearpage
and therefore:
\begin{widetext}
\begin{equation}
\begin{aligned}
\begin{split}
I_{2}(q_E^2;m_t^2,E_M^2)
=
\frac{1}{(4\pi)^2}
\int_0^\infty d\alpha \int_0^\infty d\beta\,
\frac{
\exp\!\left[
-(\alpha+\beta)m_t^2
-\dfrac{(\alpha+E_M^{-2})(\beta+E_M^{-2})}{\alpha+\beta+2E_M^{-2}}\,q_E^2
\right]
}{
\big(\alpha+\beta+2E_M^{-2}\big)^2
}.
\label{eq:workedprototypefinal}
\end{split}
\end{aligned}
\end{equation}
\end{widetext}
This equation shows in the simplest possible setting the same structural feature as Eq.\eqref{eq:IGfinal} where the entire-function regulator shifts the Schwinger quadratic form by a positive amount of order $E_M^{-2}$, makes the momentum integral Gaussian, and leaves behind a unmistakable finite parameter integral. In the local limit $E_M\to\infty$ so the regulator shifts disappear and we recover the usual local Schwinger representation.

The importance of this for the full four-loop $gg\to H$ problem is that nothing essential changes except the dimension of the quadratic form and the complexity of the polynomial numerator. The four-loop case we study replaces the scalar $A$ above by the positive-definite matrix $A(\alpha;E_M)$ of Eq.\eqref{eq:quadform} that replaces the shifted linear term by $B(\alpha;p)$, and replaces the simple prefactor $A^{-2}$ by $(\det A)^{-1/2}$, but the computational logic is the same. This is exactly why the entire-function completion we use trades the usual ultraviolet subtraction problem for direct finite parameter integration~\cite{tHooftVeltman1972,BolliniGiambiagi1972,ChetyrkinTkachov1981,Laporta2000,Kotikov1991,Remiddi1997,Henn2013}.

It is also useful to contrast this with the amplituhedron and related on-shell geometric methods, we know that the amplituhedron was developed for planar $\mathcal N=4$ super-Yang--Mills theory where they study on-shell integrands with maximal symmetry and positivity structures~\cite{ArkaniHamed:2012nw,ArkaniHamedTrnka2014}. Even the extensions beyond strict planarity still remain within the on-shell $\mathcal N=4$ framework~\cite{ArkaniHamed:2014bca}. In contrast the present problem is a nonplanar, massive, Standard Model form-factor computation with a colour-singlet Higgs insertion and with entire-function-regulated off-shell propagators. For this reason, amplituhedron-based methods do not directly furnish the object computed here, so the relevant comparison is not that positive geometry is uninteresting per se, but that it is adapted to a different class of amplitudes that may not be suitable to our universe making it less physical. Our method is instead built exactly for the regulated Standard Model quantity that enters the virtual $ggH$ form factor and is useful when we want to do real physics. This is not saying that we should not study amplituhedrons but we do need to be realistic about their physicality and relativeness to physics.

Finally, the relation to collider data must be made apparent, the quantity computed directly by Eq.~\eqref{eq:F3sum} is the raw virtual four-loop coefficient in the nonlocal theory, not yet the complete hadronic cross section. Before comparison with a conventional QCD prediction, the coupling must first be converted using Eqs.~\eqref{eq:couplingmatching} and \eqref{eq:coefficientconversion}. A direct comparison with LHC Higgs data would then additionally require combining the coupling-matched virtual correction with real-emission contributions, infrared subtraction, convolution with parton distribution functions, and the usual experimental fiducial cuts~\cite{Anastasiou2015N3LOthreshold,Mistlberger2018N3LO}. But Eq.~\eqref{eq:IGfinal} supplies the ultraviolet-finite four-loop virtual building block for that program. In the limit $E_M\to\infty$, the explicit nonlocal dependence of the graph integrals disappears, and after finite conversion to a common coupling scheme the result must reproduce the standard local-QCD form factor used in gluon-fusion phenomenology~\cite{Dawson1991,Anastasiou2015N3LOthreshold,Mistlberger2018N3LO}.

So again we note that the relation to collider data needs to be stated carefully as the quantity computed in this paper is the virtual form factor $F^{(3)}$, and not yet the full hadronic cross section needed for experimental verification. A direct comparison with LHC Higgs data we require combining the virtual correction with real-emission contributions, infrared subtraction, convolution with parton distribution functions, and the usual experimental fiducial cuts~\cite{Anastasiou2015N3LOthreshold,Mistlberger2018N3LO}. But Eq.~\eqref{eq:IGfinal} supplies a UV-finite virtual building block for exactly that program and is still very useful. In the decoupling limit of $E_M\to\infty$ the regulated result must reproduce the standard local Standard Model prediction used in gluon-fusion phenomenology~\cite{Dawson1991,Anastasiou2015N3LOthreshold,Mistlberger2018N3LO}. So the immediate phenomenological role of the current work is twofold, the first point is to recover the known local $gg\to H$ result in the large-$E_M$ limit, and the second is to parameterize any finite-$E_M$ deviation as a correction to the virtual part of the gluon-fusion cross section which can then be bounded by comparison with LHC measurements once the full hadronic prediction is assembled.

We now are able to state the key finiteness result.

\begin{theorem}[Fixed-order finiteness of four-loop $ggH$ graphs]
Assume $F$ is admissible in the sense of \eqref{eq:admissibleF} and is implemented gauge-covariantly so that no new poles are introduced~\cite{Moffat1990,Evens1991,MoffatThompsonGaugeEntire2026,LSZ1955,Cutkosky1960,ThompsonCovariance2026}. Then every raw four-loop graph contribution $I_G$ to the $gg\!\to\!H$ virtual form factor is absolutely convergent in four dimensions after Wick rotation and admits the explicit finite parameter representation \eqref{eq:IGfinal}. The conversion from the finite entire-function coupling to a conventionally normalized coupling is a separate finite matching operation and does not affect the absolute convergence of $I_G$.
\label{thm:finiteness}
\end{theorem}

After Wick rotation, each internal line contributes an entire factor $F(-\ell_{rE}^2/E_M^2)$ that is bounded by $\exp(-c\,\ell_{rE}^2/E_M^2)$ for large Euclidean momenta. The remaining integrand is a rational function times a polynomial numerator $N_G$. Because the set of line momenta $\{\ell_r\}$ spans the loop-momentum space in any connected diagram, the sum of $\ell_{rE}^2$ controls $|K|^2$ from below, so the full integrand is bounded by a polynomial times $\exp(-c'|K|^2/E_M^2)$ for some $c'>0$. So the Euclidean $16$-dimensional integral over $K$ converges absolutely. Introducing Schwinger parameters yields an overall Gaussian in $K$ and completing the square gives the closed form \eqref{eq:IGfinal}.

\section{EXTERNAL STATES AND THE LOCAL LIMIT}

To preserve the standard on-shell normalization of external gluon states, we require that $F(0)=1$, so that massless external legs are unrescaled on shell~\cite{F0Normalization}. Since $F$ is entire and introduces no additional finite-plane poles, the LSZ pole locations and physical threshold structure are unchanged~\cite{LSZ1955,Cutkosky1960,F0Normalization,ThompsonMacrocausality2026}. This statement concerns the external pole structure and residue normalization only. It does not define the finite QCD coupling and does not imply that:
\begin{equation}
\alpha_s^{(F)}(\mu,E_M)
=
\alpha_s^{\overline{\mathrm{MS}}}(\mu).
\end{equation}
The relation between these couplings is just the finite matching relation \eqref{eq:couplingmatching}. After the coupling and all other input parameters are expressed in one common wat, the regulated form factor has a well-defined local limit as $E_M\to\infty$.

This should be understood as part of the broader asymptotic-microcausality program for controlled deformations of local quantum field theory~\cite{Thompson2026AMC}.

Operationally, inclusive cross sections are obtained by combining the virtual form factor with real-emission contributions at the same order. The present letter isolates the analytically hardest building block, namely the raw four-loop virtual integrals, and makes them finite without ultraviolet subtractions, enabling direct numerical evaluation from \eqref{eq:IGfinal} for each Feynman graph $G$. A phenomenological prediction must then include the finite coupling conversion \eqref{eq:coefficientconversion} before it is expressed in terms of the standard input value $\alpha_s^{\overline{\mathrm{MS}}}(\mu)$.

\section{CONCLUDING REMARKS}

In this paper we gave an explicit analytic reduction of the four-loop N$^3$LO virtual $gg\!\to\!H$ amplitude to finite Schwinger-parameter integrals in four dimensions. The important simplification is structural where entire-function damping converts the UV-sensitive integration-by-parts or differential-equation problem into a convergent integral evaluation problem with closed-form loop-momentum integration~\cite{ChetyrkinTkachov1981,Laporta2000,Kotikov1991,Remiddi1997,Henn2013,tHooftVeltman1972,BolliniGiambiagi1972,ArkaniHamed:2012nw,ArkaniHamedTrnka2014,ArkaniHamed:2014bca}.

The ultraviolet finiteness of the four-loop graph integrals does not eliminate the ordinary freedom to choose a finite coupling-normalization prescription. We therefore defined $\alpha_s^{(F)}(\mu,E_M)$ through a gauge-invariant background-field condition and related it to $\alpha_s^{\overline{\mathrm{MS}}}(\mu)$ through finite matching. At N$^3$LO this conversion adds the lower-order terms shown in Eq.~\eqref{eq:F3matched}. So the finite graph representation and the finite coupling conversion are distinct ingredients, the former establishes ultraviolet convergence, while the latter is required to express the result in the standard phenomenological QCD scheme.

More broadly this calculation sits within an asymptotic-locality program in which sharp point-localization is replaced by a controlled quasi-local notion compatible with the nonlocal scale of the theory~\cite{Thompson2026AL}.

\section{Acknowledgments}
I would like to thank my supervisor John Moffat, my friends and colleagues Hilary Carteret and Arvin Kouroshnia for helpful discussions on non-local quantum field theory, locality, measurement in relativistic quantum theory, and many future prospects for research in this area.


\begin{thebibliography}{99}

\bibitem{F0Normalization}
E.~J.~Thompson,
``Asymptotic Microcausality to Macrocausality: Complete Quantum Electrodynamics
and the Regulated $S$-Matrix,''
In Press Annalen der Physik, (2026),
doi:10.20944/preprints202602.0207.v1.

\bibitem{LSZ1955}
H.~Lehmann, K.~Symanzik, and W.~Zimmermann,
``Zur Formulierung quantisierter Feldtheorien,''
Nuovo\ Cim.\ \textbf{1}, 205 (1955).

\bibitem{Cutkosky1960}
R.~E.~Cutkosky,
``Singularities and discontinuities of Feynman amplitudes,''
J.\ Math.\ Phys.\ \textbf{1}, 429 (1960).

\bibitem{tHooftVeltman1972}
G.~'t~Hooft and M.~Veltman,
``Regularization and renormalization of gauge fields,''
Nucl.\ Phys.\ B \textbf{44}, 189 (1972).

\bibitem{BolliniGiambiagi1972}
C.~G.~Bollini and J.~J.~Giambiagi,
``Dimensional renormalization: The number of dimensions as a regularizing parameter,''
Phys.\ Lett.\ B \textbf{40}, 566 (1972).

\bibitem{ChetyrkinTkachov1981}
K.~G.~Chetyrkin and F.~V.~Tkachov,
``Integration by parts: The algorithm to calculate $\beta$-functions in 4 loops,''
Nucl.\ Phys.\ B \textbf{192}, 159 (1981).

\bibitem{Laporta2000}
S.~Laporta,
``High-precision calculation of multiloop Feynman integrals by difference equations,''
Int.\ J.\ Mod.\ Phys.\ A \textbf{15}, 5087 (2000).

\bibitem{Kotikov1991}
A.~V.~Kotikov,
``Differential equations method: New technique for massive Feynman diagrams calculation,''
Phys.\ Lett.\ B \textbf{254}, 158 (1991).

\bibitem{Remiddi1997}
E.~Remiddi,
``Differential equations for Feynman graph amplitudes,''
Nuovo\ Cim.\ A \textbf{110}, 1435 (1997).

\bibitem{Henn2013}
J.~M.~Henn,
``Multiloop integrals in dimensional regularization made simple,''
Phys.\ Rev.\ Lett.\ \textbf{110}, 251601 (2013).

\bibitem{BognerWeinzierl2010}
C.~Bogner and S.~Weinzierl,
``Feynman graph polynomials,''
Int.\ J.\ Mod.\ Phys.\ A \textbf{25}, 2585 (2010).

\bibitem{Dawson1991}
S.~Dawson,
``Radiative corrections to Higgs boson production,''
Nucl.\ Phys.\ B \textbf{359}, 283 (1991).

\bibitem{Anastasiou2015N3LOthreshold}
C.~Anastasiou, C.~Duhr, F.~Dulat, E.~Furlan, T.~Gehrmann, F.~Herzog, and B.~Mistlberger,
``Higgs boson gluon-fusion production in QCD at three loops,''
Phys.\ Rev.\ Lett.\ \textbf{114}, 212001 (2015).

\bibitem{Mistlberger2018N3LO}
B.~Mistlberger,
``Higgs Boson Production at Hadron Colliders at N$^3$LO in QCD,''
Phys.\ Rev.\ Lett.\ \textbf{120}, 252002 (2018).

\bibitem{Baikov2009FormFactors}
P.~A.~Baikov, K.~G.~Chetyrkin, A.~V.~Smirnov, V.~A.~Smirnov, and M.~Steinhauser,
``Quark and gluon form factors to three loops,''
Phys.\ Rev.\ Lett.\ \textbf{102}, 212002 (2009).

\bibitem{Gehrmann2010eps2}
T.~Gehrmann, E.~W.~N.~Glover, T.~Huber, N.~{\c{I}}kizlerli, and C.~Studerus,
``The quark and gluon form factors to three loops in QCD through to ${\cal O}(\epsilon^2)$,''
JHEP \textbf{11}, 102 (2010).

\bibitem{LeeSmirnovSmirnovSteinhauser2022}
R.~N.~Lee, A.~V.~Smirnov, V.~A.~Smirnov, and M.~Steinhauser,
``The Higgs-gluon form factor at four loops in QCD,''
Phys.\ Rev.\ Lett.\ \textbf{128}, 212002 (2022).

\bibitem{Thompson2026AR}
E.~J.~Thompson,
``Asymptotic Renormalization: Complete Quantum Electrodynamics,''
In Press Annalen der Physik, (2026),
doi:10.20944/preprints202601.2049.v1.

\bibitem{Moffat2025EWSB}
J.~W.~Moffat,
``Electroweak Symmetry Breaking,''
arXiv:2503.11548 [hep-ph] (2025).

\bibitem{Moffat1990}
J.~W.~Moffat, “Finite nonlocal gauge field theory,” arXiv:1407.2086 Phys. Rev. D 41, 1177 – Published 15 February, (1990).

\bibitem{Evens1991}
D.~Evens, J.~ W. ~Moffat, G.~Kleppe and R.~P.~Woodard, "Nonlocal regularizations of gauge theories,"
Phys. Rev. D 43, 499 – Published 15 January, (1991).

\bibitem{MoffatThompsonGaugeEntire2026}
J. W. Moffat and E. J. Thompson,
``On Gauge-Invariant Entire-Function Regulators and UV Finiteness
in NonLocal Quantum Field Theory,''
Ann. Phys. (Berlin) \textbf{538}, e70207 (2026),
arXiv:2511.11756 [hep-th],
doi:10.1002/andp.70207.

\bibitem{Abbott1982}
L.~F.~Abbott,
``Introduction to the Background Field Method,''
Acta\ Phys.\ Polon.\ B \textbf{13}, 33 (1982).

\bibitem{ModestoRachwal2015}
L.~Modesto and L.~Rachwa{\l},
``Universally finite gravitational and gauge theories,''
Nucl.\ Phys.\ B \textbf{900}, 147 (2015).

\bibitem{ModestoPivaRachwal2016}
L.~Modesto, M.~Piva, and L.~Rachwa{\l},
``Finite quantum gauge theories,''
Phys.\ Rev.\ D \textbf{94}, 025021 (2016).

\bibitem{Schwinger1951}
J.~Schwinger,
``On Gauge Invariance and Vacuum Polarization,''
Phys.\ Rev.\ \textbf{82}, 664 (1951).

\bibitem{ArkaniHamedTrnka2014}
N.~Arkani-Hamed and J.~Trnka,
``The Amplituhedron,''
JHEP \textbf{10}, 030 (2014).

\bibitem{ArkaniHamed:2012nw}
N.~Arkani-Hamed, J.~L.~Bourjaily, F.~Cachazo,
A.~B.~Goncharov, A.~Postnikov, and J.~Trnka,
``Scattering Amplitudes and the Positive Grassmannian,''
arXiv:1212.5605 [hep-th].

\bibitem{ArkaniHamed:2014bca}
N.~Arkani-Hamed, J.~L.~Bourjaily, F.~Cachazo,
A.~Postnikov, and J.~Trnka,
``On-Shell Structures of MHV Amplitudes Beyond the Planar Limit,''
JHEP \textbf{06} (2015) 179,
arXiv:1412.8475 [hep-th].

\bibitem{Thompson2026AMC}
E.~J.~Thompson,
``Asymptotic Microcausality; Deformations of Local Quantum Field Theory,'' In Press Annalen der Physik, (2026),
doi:10.20944/preprints202602.0207.v1.

\bibitem{Thompson2026AL}
E.~J.~Thompson,
``Asymptotic Locality Rectifying Newton--Wigner and Foldy--Wouthuysen Localization,''
In Press Annalen der Physik, (2026).

\bibitem{ThompsonLocalization2026}
E. J. Thompson,
``On the Meaning of Localization in Non-Local Quantum Field Theory and On the Limits of a Space-Time Description and the Physical Meaning of Phase Space in a Nonlocal Continuum,''
Ann. Phys. (Berlin) \textbf{538}, e70233 (2026),
arXiv:2606.27387v2.

\bibitem{ThompsonCovariance2026}
E. J. Thompson,
``On Covariance for Entire-Function Deformations of Relativistic
Field Theories,''
Preprints (2026),
doi:10.20944/preprints202604.0706.v1.

\end{thebibliography}
\end{document}